\documentclass[preprint2]{aastex}

\usepackage{multicol}

\def\lsim   {\stackrel{<}{\sim}}

\def\muK    {\mu K}
\slugcomment{ }

\shorttitle{ }
\shortauthors{ }

\begin{document}
\title{Search for Cosmic Strings in CMB  Anisotropies \\}

\author{E. Jeong}
\affil{Department of Physics, University of California, Berkeley, CA 94720}
\email{ehjeong@socrates.berkeley.edu}
\author{G. F. Smoot}
\affil{Department of Physics, University of California, Berkeley, CA 94720}
\email{gfsmoot@lbl.gov}

\begin{abstract}
We have searched the 1st-year WMAP W-Band CMB anisotropy map for evidence of 
cosmic strings. We have set a limit of $\delta = 8 \pi G \mu / c^2 < 8.2 
\times 10^{-6}$ at 95\% CL for statistical search for a significant number of 
strings in the map. We also have set a limit using the uniform distribution of 
strings model in the WMAP data with $\delta = 8 \pi G \mu / c^2 <  7.34 \times 
10^{-5}$ at 95\% CL. And the pattern search technique we developed here set 
a limit $\delta = 8 \pi G \mu / c^2 <  1.54 \times 10^{-5}$ at 95\% CL.
\end{abstract}


\section{Introduction}
Current theories of particle physics predict that topological defects would 
almost certainly be formed during the early evolution of the universe [1]. 
Just as liquids turn to solids when the temperature drops, 
so the interactions between elementary particles run through distinct phases as
the typical energy of those particles decreases with the expansion of the
universe. When conditions favor the appearance of a new phase, the new phase
crops up in many places at the same time, and when separate regions of the new
phase run into each other, topological defects are the result.
The detection of defects in the modern universe would provide precious
information on events in the earliest moments after the Big Bang.
Their absence, on the other hand, would force a major revision of current
physics theories. The least of which would be to have the phase transitions be
second or higher order.

The potential role of cosmic topological defects in the evolution
of our universe has interested the astrophysical community for many years.
Combined theoretical and experimental work has led to the development of
observational signatures in a variety of diverse data sets.
These have, on one hand removed one major motivation of the primary role in 
structure formation and on the other raised new motivations.
These have also narrowed the range of allowable topological defects
so that most viable models are likely to produce some variety of cosmic strings.

CMB anisotropy power spectrum observations rule out topological defects as the 
primary source of structure in the universe [2,3]. Observations strongly favor 
adiabatic random fluctuations from something like inflation rather than the 
structure formation from topological defects (e.g. cosmic strings).
These CMB results, while confirming inflation, do not rule out lower level 
contributions from topological defects. Interest in cosmic strings has, 
in fact, been renewed with recent theoretical work on hybrid inflation, 
D-Brane inflation and SUSY GUTS. The idea that inflationary cosmology might 
lead to cosmic string production is not new; however, it has received new 
impetus from the brane world scenario suggested by superstring theory.
A seemingly unavoidable outcome of brane inflation is the production
of a network of cosmic strings [4], whose effects on cosmological observables 
range from negligible to substantial, depend on the specific brane 
inflationary scenario [5]. A number of theory papers [12,13,14], anticipating 
that searches from cosmic strings will be negative and set significant limits, 
have begun developing modified theories so as not to produce them. At 
minimum these modified models must introduce a new field (also warping brane).

Since a moving string would produce a steplike discontinuity in the CMB,
it will cause temperature distribution deviate from Gaussian.
We may be able to detect non-Gaussian aspects of temperature distribution
provided sufficient resolution.
We search for strings in two ways: statistical and pattern-discovery methods.
Statistical analysis determines how much the distribution of temperature
fluctuation deviates from Gaussian distribution or what fraction of the
fluctuations might be due to strings.
In the second approach, we can search for cosmic strings directly from
temperature map by their distinctive pattern of anisotropy. The latter approach
is much easier and more straight forward when CMB signal is not contaminated
seriously. These methods are distinctly different than a simple fitting to 
angular power spectrum [4].
\section{Effect of Cosmic Strings on CMB}
\subsection{Signal from a moving Cosmic String}
Consider a cosmic string with mass per unit length $\mu$, velocity 
$\vec\beta $ and direction $\hat s$ both of which are perpendicular to the 
line of sight and is backlighted by a uniform blackbody radiation background of
temperature $T$. Due to its angular defect $\delta = 8\pi G\mu /c^2$, there is 
a Doppler effect of one side of the string relative to the other which causes a
temperature step across the string [8,9] 
\begin{equation}
\label{a}\frac{\delta T}{T} = 8 \pi G \mu \gamma \beta /c^2 .
\end{equation}
where $\gamma = 1/\sqrt{1-\beta^2}$. This expression was generalized for 
arbitrary angles between the string direction $\hat s$, its velocity 
$\vec \beta$, and the line of sight $\hat n$ [6,7]
\begin{equation}
\label{b}\frac{\delta T}{T} = 8 \pi G \mu \gamma \beta\hat n \cdot ( \hat \beta
\times \hat s) /c^2 .
\end{equation}
\subsubsection{Probability distributions for relevant parameters}
\begin{figure}[h]
\epsscale{1.0}
\plotone{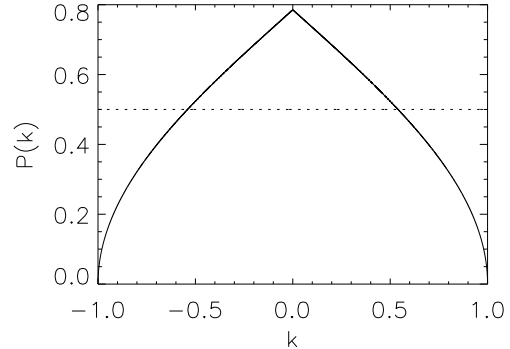}
\caption{Probability distributions for temperature steps in equation
(\ref{b}). The solid curve represents the probability with both string 
direction $\hat s$ and string velocity $\vec \beta$ having random directions 
while the flat dotted line assumes $\hat\beta \bot \hat s$ ($\sin\phi =1$) but 
the direction of $\hat\beta \times \hat s$ is random relative to the line of 
sight ($\cos\theta$ random).}
\end{figure}
$\delta T/T$ in equation (\ref {b}) is determined by three factors $G\mu /c^2$,
$\gamma \beta$ and  $\hat n \cdot ( \hat \beta \times \hat s)$ which arise from
 different sources. $G\mu /c^2$ is related to symmetry breaking scale,
we can assume $\gamma \beta \sim 1$ except near cusps where one can have
$\gamma \gg 1$ and  $\hat n \cdot ( \hat \beta \times \hat s)$ depends on the
combination of relative angles among $\hat \beta$, $\hat s$ and $\hat n$.
We can also assume that $\hat \beta \times \hat s $ is random in direction to the line of sight $\hat n$ in 3-D space.
If we denote $\hat \beta \times \hat s$ as $\hat u \,\sin \phi$, then
\begin{equation}
\label{c}\hat n \cdot ( \hat \beta \times \hat s)=\hat n \cdot \hat u \,\sin \phi =\cos \theta\,\sin\phi\; .
\end{equation}
The infinitesimal probability that $\cos \theta$ has an arbitrary value is
$2\pi\sin \theta d\theta /4\pi = \frac{1}{2}d\cos\theta$ and thus its
probability distribution is uniform in $\cos\theta$.
If the string velocity $\vec\beta$ and its direction $\hat n$ are uncorrelated,
the probability distribution for $\sin \phi$ is $y$ is $y/\sqrt{1-y^2}$.
Then, the probability distribution for $k\equiv\hat n \cdot (\hat\beta\times\hat s)$ being an arbitrary value $-1< k <1$ becomes, substituting $\cos \theta =x$,\begin{eqnarray}
P(k) &=& \int^{1}_{-1}\frac{1}{2}dx\int^{1}_{0}\frac{y}{\sqrt{1-y^2}}\delta (k-x\,y)dy\nonumber\\
&=& \frac{1}{2}\int^{1}_{|k|}\frac{k\,dx}{x\sqrt{x^2-k^2}}\nonumber\\
\label{d} &=& \frac{1}{2}\cos^{-1}|k|\, .
\end{eqnarray}
Often string velocity and direction will be perpendicular to each other so that
$\sin\phi \simeq 1$ and in this case the probability reduces to
$P(k)=\frac{1}{2}$. These probability distributions are plotted in Figure 1.

In a matter dominated universe the projected angular length of string in the 
redshift interval [$z_1,  ~ z_2$] scales as $\sqrt{z_1} - \sqrt{z_2}$ [10]. 
The standard cosmological model ($\Omega_{DE}\sim 0.7,\:\Omega_m\sim 0.3$) 
gives $z_{ls}\sim 1100$ [10]. The apparent angular size of the horizon at the 
CMB last scattering redshift $z_{ls}$ is
\begin{equation}
\label{f}\theta_H = \frac{1}{\sqrt{z_{ls}}} radians = 1.8^\circ \left( \frac{1000}{z_{ls}} \right)^{1/2}\sim 1.7^\circ .
\end{equation}
The average distance between strings is roughly $d_H/3$.
Thus the typical angular distance between the discontinuities
on the sky to be of the order of $\theta_H$.
The rough expected magnitude of the jumps in temperature are of order
$\delta T / T \sim 13 G \mu /c^2~{\rm to} ~ 19 G \mu /c^2$.
With angular resolution of $2^{\prime}$ (2 arcmin) there could be sharp
jumps up to about $\delta T / T \sim 40 G \mu /c^2$.
(i.e. peak range of temperature steps in a distribution of possible steps.)
If the angular resolution is poorer, then the blurring effectively smooths the
steps that the maximum range of steps are at a somewhat smaller level.
\subsection{Statistical Fluctuations}
\begin{figure}[t]
\epsscale{1.0}
\plotone{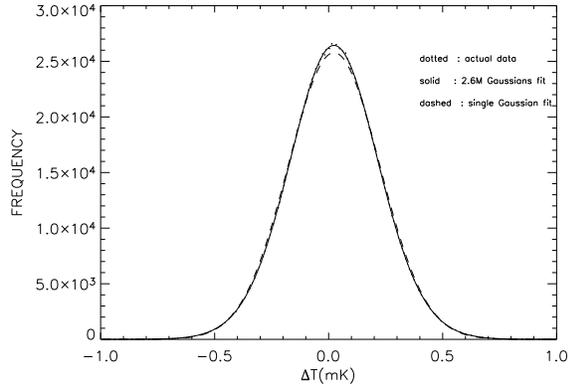}
\caption{{ Temperature Distribution of full sky from WMAP.
Pixels with $|\Delta T| > 1mK$ or $|$Galactic latitude$|<10^\circ$ are excluded
because these signals are mainly due to the Galaxy or bright sources.
The dashed curve is the best-fitted Gaussian with $\mu =23.6\mu K$ and
$\sigma =201.25\mu K$. $\chi^2 /DOF$ for the Gaussian fit is 4085.87/398 and
the solid curve represents 2.6M-Gaussians fit with $\mu =23.6\mu K$,
$\sigma_0 = 6682.22\mu K$ and $\sigma_{CMB}=80.25\mu K$. $\chi^2 /DOF$
for 2.6M-Gaussians fit is 435.76/398.}}
\end{figure}
The signal in a microwave sky map will be made of several components:
noise from the instrument, foreground signals (expected to be small away from
the galactic plane and with strong sources punched out), CMB fluctuations from
adiabatic (or appropriate) fluctuations, and potential signals from the strings.The signal observed at any pixel is then
\begin{equation}\label{g}
T_{pixel}=Noise_{pixel}+T_{foreground}+\delta T_{CMB}+\delta T_{string}.
\end{equation}
Since the nature of signal is the superposition of random Gaussian signal
(noise, $\delta T_{CMB}$) and non-Gaussian contribution ($\delta T_{string}$),
there are two basic forms of distribution functions (foreground will be removed
from the beginning). 

Here and later on in this paper, we use two identical variables $T$ and 
$\Delta T$ confusingly,  i.e, $T=\Delta T=T_{absolute}-T_0(2.73K)$.\\
(1){\it Gaussian signal}\\
Major portion of CMB signal obeys Gaussian distribution,
\begin{eqnarray}
\label{j1}f_{1}(T)&=&\frac{1}{\sigma\sqrt{2\pi}}e^{-\frac{1}{2}{\left(\frac{T-\mu}{\sigma} \right)}^2}\\
\sigma^2 &=& \sigma^2_0/n_i+\sigma^2_{CMB}\nonumber
\end{eqnarray}
where $\sigma_0$ and $\sigma_{CMB}$ are variances from noise and CMB
fluctuations each, $n_i$ is the number of observations for $i_{\mathrm{th}}$ 
pixel. We make use of the information that the noise is Gaussian (actually 
Gaussian per observation and thus variance inversely proportional to the 
number of observations per pixel) and the intrinsic CMB fluctuations are 
gaussianly distributed.\\
(2){\it Non-Gaussian signal from strings}\\
When a straight moving string is added to a region, we can expect a temperature
distribution that is the sum of two Gaussians rather than single Gaussian
because of the blue- and redshift by a transversely moving string. Then the
probability distribution in equation (\ref{j1}) is modified to
\begin{equation}
\label{n}f_{2}(T)=\frac{1}{\sigma\sqrt{2\pi}}\left[p\, e^{-\frac{1}{2}{\left(\frac{T-\mu-\delta T/2}{\sigma} \right)}^2} + q\, e^{-\frac{1}{2}{\left(\frac{T-\mu+\delta T/2}{\sigma} \right)}^2} \right]
\end{equation}
where $p=N_{blue}/N_{total}$, the ratio blueshifted pixels to total pixels, 
$q=1-p$ and $\delta T/2$ is half of the effective height of step across the 
string as given in equation (\ref{b}). 
If we denote $\nu $ and $\sigma_{obs} $ as mean and
standard deviation calculated from observation, they are related to $\mu $ and
$\sigma $ as follows
\begin{eqnarray}
\nu &=& \mu +p\,\delta T/2 +(1-p)(-\delta T/2)\nonumber\\
\label{o}    &=& \mu +(2p-1)\delta T/2\\
\sigma_{obs}^{2} &=& <(T-\nu )^2>\nonumber\\
  &=&\int^{\infty}_{-\infty} (T-\nu)^2 f_{2}(T)\,dT\nonumber\\
\label{p}  &=&\sigma^2 +4p(1-p)(\delta T/2)^2\; .
\end{eqnarray}
Here we used an approximation on the integration interval as
$-1mK < T < 1mK \longrightarrow -\infty < T < \infty $. It is a good
approximation because, $\sigma \sim 200\mu K$ and thus $1mK \sim 5\sigma$,
the variance integral in equation (\ref{p}) over $[-5\sigma,5\sigma]$ covers
99.9985\% of that over $[-\infty,\infty]$. This approximation is also implied
in equation (\ref{r}).
\begin{table}[t]
\begin{center}
\begin{tabular}{ccccc}
\tableline\tableline
Signal  & Limit $\sigma_{string}$ & $\delta T/T$ & Limit $G\mu /c^2$ & 
GUT symmetry breaking\\
  & $\mu$K   &  $\delta = 8\pi G\mu \gamma \beta / c^2 $& $\times 10^6$ & 
scale $\eta_{SB}$ ($\times 10^{16}$GeV)\\
\tableline
Total  & 201.25 & 7.37 $\times 10^{-5}$ & 2.93 & 2.05\\
Total - Noise 6.7mK/obs & 80.26 & 2.94$ \times 10^{-5}$ & 1.17 & 1.30\\
Total - Noise - Adiabatic CMB & 22.49 & 0.82 $\times 10^{-5}$ & 0.33 & 0.69\\
Uniform Distribution of Strings & 57.82 & 7.34$\times 10^{-5}$ & 2.92 & 2.05\\
Pattern Search & N/A & 1.54$\times 10^{-5}$ & 0.61 & 0.94\\
\tableline
\end{tabular}
\parbox{16cm}
{\caption{ Variance Limits on Cosmic Strings. All the values presented are at 
95\% CL. The model 'Uniform Distribution of Strings' on the fourth row is the 
distribution given in the equation (\ref{q}). The GUT symmetry breaking scales,
$\eta_{SB}$, in the last column are calculated using the relation 
$\eta_{SB}\sim m_{pl}\sqrt{G\mu/c^2}$ with $m_{pl}=1.2\times 10^{19}$GeV.}}
\end{center}
\end{table}
Temperature distribution of sky may be explained with appropriate combinations
of $f_{1}(T)$ and $f_{2}(T)$. All the models introduced in the following
sections are based on these two primary patterns of temperature distribution.
\subsubsection{Variance: Quadratic Estimator Test for String Contribution}
Assuming that the signals are all statistically independent,
we can find the variance in the map
\begin{equation}
 \label{h}\sigma^2_S = \sigma^2_{noise} + \sigma^2_{foreground} + \sigma^2_{CMB} + \sigma^2_{string}
\end{equation}
\\
\\
\\
\\
\\
\\
\\
\\
\\
\\
\\
\\
\\
We remove the significant foreground contribution by dropping pixels with
$|$Galactic latitude$|<10^\circ$ or $|\Delta T| > 1mK$. 
The first condition is to exclude galactic area where non-CMB signal is 
dominant. We lose 217 more pixels by imposing the condition 
$|\Delta T| < 1mK$, 4 pixels are less than $-1mK$ and 213 pixels are greater 
than $1mK$. Since Gaussian tail probability allows $\sim 1.7$ pixels in 
$|\Delta T|>1mK$ region and they mostly form clusters, we can drop those 
pixels as they are from unusual bright sources. 
We introduce a temperature distribution based on the Gaussianity of 
signal for each pixel in WMAP 1st-year data and calculate $\sigma_{CMB}$ which 
contains all the contributions except instrumental noise,
\begin{eqnarray}
\label{h1}f_{2.6M}(T)\!&=&\!\frac{1}{N}\sum_{i}\frac{1}{\sigma_i\sqrt{2\pi}}e^{-\frac{1}{2}{\left(\frac{T-\mu}{\sigma_i} \right)}^2}\nonumber\\
\sigma^2_{i} \!&=&\!\frac{\sigma_{0}^2}{n_{i}}+\sigma^2_{CMB}
\end{eqnarray}
where $N$ is the total number of pixels, 2,598,695, $i$ is pixel number with
maximum resolution of WMAP 1st-year data, $\sigma_0^2$ is variance due to noise
per measuremet and $n_i$ is effective number of measurements for $i_{th}$ pixel. The distribution $f_{2.6M}(T)$ in equation (\ref{h1}) has minimum $\chi^2$ for
$\sigma_{CMB} = 80.25^{+0.05}_{-1.65} \mu K$ and
$\sigma_0 = 6682.22^{+21.78}_{-3.22} \mu K$ and thus
$\sigma_{S} = 200.68^{+0.57}_{-0.74} \mu K$ each with 95$\%$ CL.
This observation in turn allows an upper limit of order
$\sigma_{string} \lsim 200 \mu K$ and thus providing a limit of
$8\pi G \mu /c^2 \lsim  200 \mu K /2.73 K = 73.5 \times 10^{-6}$.

We obtain a better upper limit, if we know the mean contribution of 
instrumental noise and other signals.
According to the WMAP 1st-year data release, the sum over pixels of (number of
measurements per pixel)$^{-1}$ for 2,598,695 pixels included here is 1968.96.
Thus the effective variance due to noise is
\begin{eqnarray}
 \sigma^2_{noise} \!&=&\! (\sigma^2_0 \sum_i \frac{1}{n_i})/N\nonumber\\
\!&=&\! (6682.22^{+21.78}_{-3.22})^2\frac{ 1968.96}{2598695}\nonumber\\
\!&=&\! (183.93^{+0.6}_{-0.08}\: \mu K)^2 ,\; 95\%\; {\mathrm {CL}}\label{sp0}
\end{eqnarray}
where $n_i$ is the effective number of measurements for $i_{th}$ pixel.

\begin{figure}[t]
\epsscale{1.0}
\plotone{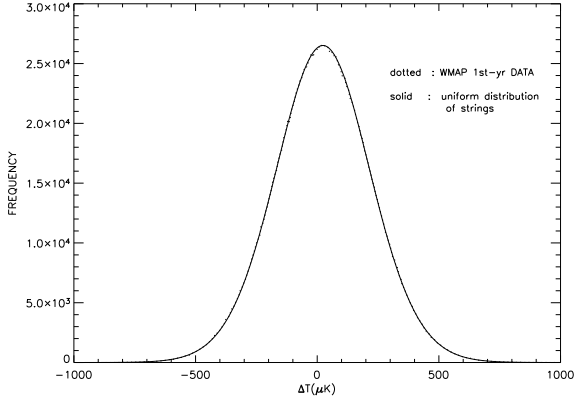}
\caption{Temperature distribution of full sky excluding the pixels with
$|$Galactic latitude$| < 10^{\circ}$ or $|\Delta T| > 1mK$.
Solid curve represents the best-fitted curve of uniformly distributed strings
model and the dotted curve shows actual WMAP 1st-year data. }
\end{figure}
Under the assumption that the signals other than noise and from strings
are negligible,
\begin{equation}
\label{sp1}\sigma^2_{string}\leq\sigma^2_S-\sigma^2_{noise}
=(80.25^{+0.05}_{-1.65}\: \mu K)^2 ,\; 95\%\; {\mathrm {CL}}.
\end{equation}
And we are able, by taking the instrument noise into account, to set a lower
upper limit.
\begin{figure}[t]
\epsscale{1.0}
\plotone{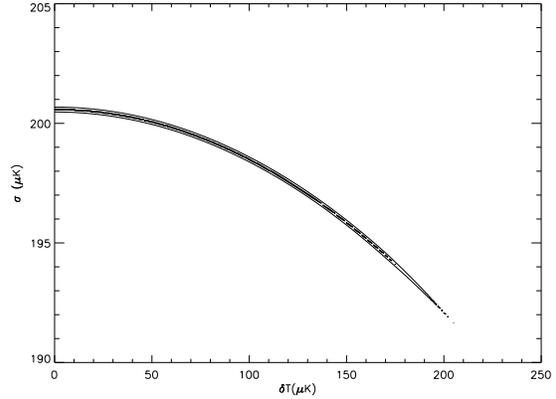}
\caption{$\chi^2$ distribution for the uniformly distributed strings model.
$\chi^2_{min}=435.43$ with 398 degrees of freedom. Parameters span
$0<\delta T<200.3\,\mu K$ and $192.1\,\mu K <\sigma<200.8\,\mu K $ at 95\% CL.
Here $\sigma^2$ represents the total variance except string contribution,
$\sigma^2=\frac{\sigma^2_0}{N}\sum_{i}1/{n_i}+\sigma^{2}_{CMB}$.
The outer contour is for 95\% CL which is almost overlapped with the 68\% CL
contour in vertical direction.}
\end{figure}

Taking into account the CMB anisotropy spectrum is characteristic of 
anisotropies that arise from stochastic, Gaussianly-distributed, adiabatic
primordial perturbations. We decompose the variance $\sigma^2_{CMB}$ computed
in equation (\ref{sp1}) into two contributions, one from adiabatic fluctuation
and one from strings, $\sigma^2_{CMB}=\sigma^2_{adiabatic}+\sigma^2_{string}$. 
Since $\sigma^2_{CMB}=\sum_l (2l+1)C_l B_l^2/4\pi$ and string contribution in 
$C_l$ is $\sim 10\%$ [4] where $B_l$ is beam filter function, 
we estimate the upper limit of 
$\sigma^2_{string}\sim\sigma^2_{CMB}/10=(25.38^{+0.50}_{-1.04}\:\mu K)^2$, thus
\begin{equation}
\label{j} 0<\sigma_{string}<25.38^{+0.50}_{-1.04}\mu K ,\; 95\%\; {\mathrm{CL}}.
\end{equation}
We can also calculate $\sigma_{adiabatic}^2$ directly from the relation between
$\sigma_{CMB}^2$ and the power spectrum coefficients $C_l$. Using the $C_l$'s
of best-fitted modeled cosmological parameters ($\Omega_{\Lambda}=0.73,\;
\Omega_{b}=0.046,\; \Omega_{cdm}=0.224)$, $B_l$'s from WMAP 1st-year data and
convolving with finite pixel size, we have
\begin{eqnarray}
\sigma_{CMB}^2 \!\!&=&\!\!\frac{1}{4\pi}\sum_{l}(2l+1)C_l B_l^2\nonumber\\
\!\!&=&\!\!(79.085\pm 0.376\mu K)^2 ,\; 95\%\; {\mathrm{CL}}.\label{j3}
\end{eqnarray}
for beam and finite pixel size.
Assuming that $\sigma_{CMB}^2$ in equation (\ref{j3}) is solely from adiabatic
fluctuation, we have even lower upper limit on $\sigma_{string}$,
\begin{eqnarray}
&\sigma_{string}^2 &\!\le \:\sigma^2_{S}-\sigma^2_{noise}-\sigma^2_{adiabatic}\nonumber\\
&=&\!\!\!\!\! (201.25\mu K)^2-(183.85\mu K)^2-(78.709\mu K)^2\nonumber\\
&=&\!\!\!\!\! (22.49\mu K)^2 ,\; 95\%\; {\mathrm{CL}}.\label{j2}
\end{eqnarray}
Both of these last two are model dependant as one could refit all cosmological 
parameters to include a small addition of string. Presumably that is what was 
done in [4].
\subsubsection{Statistical Test for Temperature Step Expected from Random Strings}
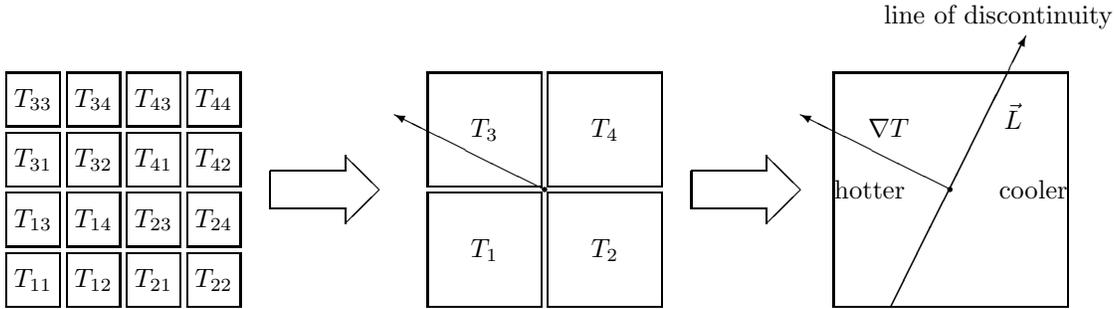
\begin{figure}[t]
\setlength{\unitlength}{1cm}
\begin{center}
\begin{picture}(10,4)
\put(0.0,0.5){\framebox(0.7,0.7){$T_{11}$}}
\put(0.8,0.5){\framebox(0.7,0.7){$T_{12}$}}
\put(1.6,0.5){\framebox(0.7,0.7){$T_{21}$}}
\put(2.4,0.5){\framebox(0.7,0.7){$T_{22}$}}
                                                                                
\put(0.0,1.3){\framebox(0.7,0.7){$T_{13}$}}
\put(0.8,1.3){\framebox(0.7,0.7){$T_{14}$}}
\put(1.6,1.3){\framebox(0.7,0.7){$T_{23}$}}
\put(2.4,1.3){\framebox(0.7,0.7){$T_{24}$}}
                                                                                
\put(0.0,2.1){\framebox(0.7,0.7){$T_{31}$}}
\put(0.8,2.1){\framebox(0.7,0.7){$T_{32}$}}
\put(1.6,2.1){\framebox(0.7,0.7){$T_{41}$}}
\put(2.4,2.1){\framebox(0.7,0.7){$T_{42}$}}
                                                                                
\put(0.0,2.9){\framebox(0.7,0.7){$T_{33}$}}
\put(0.8,2.9){\framebox(0.7,0.7){$T_{34}$}}
\put(1.6,2.9){\framebox(0.7,0.7){$T_{43}$}}
\put(2.4,2.9){\framebox(0.7,0.7){$T_{44}$}}
                                                                                
\put(3.5,1.8){\line(0,1){0.5}}
\multiput(3.5,1.8)(0,0.5){2}{\line(1,0){1}}
\multiput(4.5,1.6)(0,0.7){2}{\line(0,1){0.2}}
\put(4.5,1.6){\line(1,1){0.45}}
\put(4.5,2.5){\line(1,-1){0.45}}
                                                                                
\put(5.6,0.5){\framebox(1.5,1.5){$T_1$}}
\put(7.2,0.5){\framebox(1.5,1.5){$T_2$}}
\put(5.6,2.1){\framebox(1.5,1.5){$T_3$}}
\put(7.2,2.1){\framebox(1.5,1.5){$T_4$}}
                                                                                
\put(7.15,2.05){\vector(-2,1){2}}
\put(7.15,2.05){\circle*{0.07}}
                                                                                
\put(9.1,1.8){\line(0,1){0.5}}
\multiput(9.1,1.8)(0,0.5){2}{\line(1,0){1}}
\multiput(10.1,1.6)(0,0.7){2}{\line(0,1){0.2}}
\put(10.1,1.6){\line(1,1){0.45}}
\put(10.1,2.5){\line(1,-1){0.45}}
                                                                                
\put(11,0.5){\framebox(3.1,3.1)[s]{hotter  cooler}}
\put(11.75,0.5){\line(1,2){1.55}}
\put(12.55,2.05){\vector(-2,1){2.0}}
\put(12.54,2.05){\circle*{0.07}}
\put(11.75,0.5){\vector(1,2){1.8}}
\put(13.5,3.2){\makebox(0,0)[tr]{$\vec L $}}
\put(12.0,3.0){\makebox(0,0)[tr]{$\nabla T$}}
\put(14.7,4.5){\makebox(0,0)[tr]{line of discontinuity}}
\end{picture}
\parbox{13.5cm}{
\caption{ Generating the vector field $\vec L$. Pixels left demoted to
 middle pixels, middle $\rightarrow$ right: defining gradient of temperature
$\nabla T$ and $\vec L$ which lies along temperature step.}}
\end{center}
\end{figure}
For the anticipated distribution of strings, one would expect a random
distribution of temperature steps which has roughly equal probability between
the plus and minus the maximum temperature step amplitude.
We also fitted to a distribution that was a Gaussian for the other signals convolved with a uniform distribution of temperature steps.
Let ${\delta T_0} = 8\pi G\mu\gamma \beta /c^2 T_0$ be the
characteristic value for a string, then the actual effect of temperature step
left on the measured CMB is $k{\delta T_0}/2 $ where
$-1<k=\cos \theta\,\sin \phi <1$ from the equation (\ref{d}) and $T_0 =2.73 K$.
Assuming $\sin\phi\sim 1$ $i.e$ direction of a string and its velocity are most
likely perpendicular, we have $P(k)\sim \frac{1}{2}$.
Then, we obtain the distribution function of temperature
\begin{equation}\label{q}
f_{T}(T) = \frac{1}{N}\sum_{i}\int^{1}_{-1}f_{2}^{(i)}(T,k{\delta T_0}/{2},\sigma)dk
\end{equation}
where $f_{2}^{(i)}(T,k{\delta T_0}/{2},\sigma)$ is the two-Gaussians form 
with $p=0.5$ and $\delta T = k{\delta T_0}$ of $i$-th Gaussian in equation (\ref{h1}).
Then, the apparent variance $\sigma_{uniform}^2$ for this distribution becomes,
in terms of $\sigma $ and ${\delta T_0}$,
\begin{eqnarray}
\sigma_{uniform}^2 \!\!&=&\!\!<(T-\mu)^2>\nonumber\\
\!\!&=&\!\! \int^{\infty}_{-\infty}(T-\mu)^2 f_{T}(T)dT\nonumber\\
\label{r}\!\!&=&\!\! \frac{\sigma^{2}_{0}}{N}\sum_{i}\frac{1}{n_{i}}+\sigma^{2}_{CMB}+\frac{\delta T_{0}^2}{12}
\end{eqnarray}
where the first 2 terms are directly from the equation (\ref{h1}) and the last 
term is the contribution 
\\
\\
\\
\\
\\
\\
\\
\\
\\
\\
\\
\\
\\
\\
\\
\\
from strings. From the $\chi^2$ distribution in Figure 4, we have
$192.1\,\mu K<\sigma  <200.8 \,\mu K$ and $0<\delta T_{0}<200.3\,\mu K$ at
$95\%$ CL. Thus, temperature variation by a moving string can be limited
to, in terms of deficit angle,
\begin{equation}
\label{r1}0 <\frac{\delta T_0}{T_0}=\frac{8\pi G\mu}{c^2} < 7.34\times 10^{-5}\; ,\; 95\% {\mathrm{CL}}.
\end{equation}
\subsection{Search for Temperature Steps}
One can directly search for temperature steps from the CMB sky map by their
topological configuration or pattern. We define an algorithm to search for CMB
temperature steps produced by cosmic strings in a background of CMB adiabatic 
fluctuations and (at present even more dominant) receiver noise. (For the WMAP 
1st-year data release the signal-to-noise ratio for an average W-band pixel is 
about 0.5.) As a smoothing process, we demote $2^{2n}$ pixels of maximum 
resolution to one pixel with representative temperature value (average) 
assigned. This demoted pixel can be obtained by repeating demotion process 
shown in Figure 5 until we reach desired value of signal/noise. Since 
$\sigma_{S} \sim 201\mu K $ and $\sigma_{noise} \sim 184\mu K $ for the 
WMAP 1st-year data, it is reasonable to take 16 pixels demotion or more 
because we have $\sigma_{S}^{\left(16\right)} \sim 104\mu K$ and 
$\sigma_{noise}^{\left(16\right)}=\sigma_{noise}/\sqrt{16}\sim 46\mu K$, 
thus the contribution from noise becomes subdominant. 
Then a vector can be derived out of 4 neighboring demoted pixels. 
$\nabla T$ is defined as
\begin{eqnarray}
\label{t}(\nabla T)_x&=&\frac{1}{2}\left( T_{2}+T_{4}-T_{1}-T_{3}\right)\\
\label{u}(\nabla T)_y&=&\frac{1}{2}\left( T_{3}+T_{4}-T_{1}-T_{2}\right)\; .
\end{eqnarray}
\begin{figure}[t]
\begin{center}
\epsscale{2.5}
\plotone{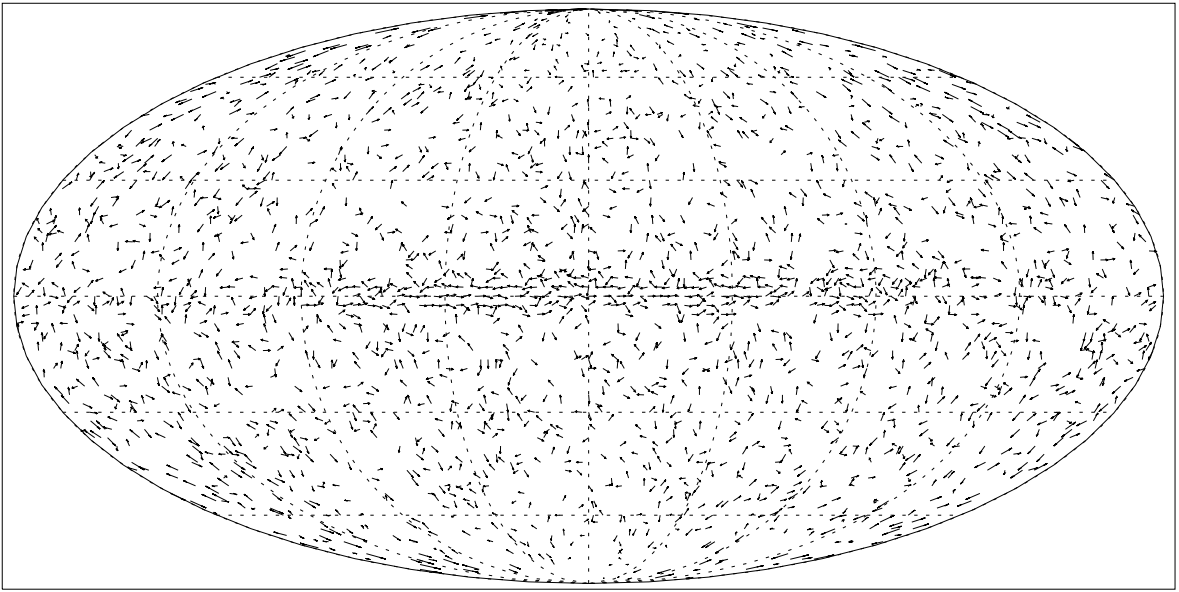}
\parbox{15cm}{
\caption{$\vec L$ field for the full sky with 64-pixel demotion.
The center of the map is ($\theta = 0^{\circ}$, $\phi = 0^{\circ}$) in Galactic
coordinates. Each arrow represents local temperature step and the temperature
steps greater than $100\mu K$ are set to $100\mu K$. The pattern along the
equator is due to the Galaxy.}}
\end{center}
\end{figure}
We build up a vector field for the full sky which contains 
$12\times 2^{16-2n}$ $\nabla T$'s where $2^{2n}$ is the demotion factor.
For each $\nabla T$, the perpendicular direction to $\nabla T$ 
represents the direction of step discontinuity. So if there is a consistent 
elongated line of discontinuity, we may interpret it as the signal due to 
moving string. After covering full sky with $\nabla T$-field, we rotate 
the vector field $-\pi/2$ to define $\vec L$-field so that each arrow lies 
along the local isothermal line with lefthand side of arrow being higher 
temperature. Figure 6 shows the $\vec L$-field map for the entire sky. The 
length of each arrow represents the magnitude of temperature gradient on that 
point and any temperature step greater than $100\mu K$ is set to $100\mu K$ to 
make steps visible which are less than $100 \mu K$. It shows a long coherent 
structure along the galactic plane, this is because there is a steep 
temperature rise approaching the galactic plane. 
\\
\\
\\
\\
\\
\\
\\
\\
\\
\\
\\
\\
\\
\\
\\
\\
\\
\\
\\
\\
\\
\\
\\
\\
\\
\\
\\
\\
\\
\\
\\
\subsection{Evaluating String Pattern}
We define the connectedness of two neighboring temperature steps with two
smoothness conditions for the heights of steps and the curve that links
neighboring $\vec L$'s.\\
(1) {\it Component Definition of Connectedness}\\
We assume that the temperature distribution of WMAP 1st-year data is
approximately Gaussian (Figure 2), $f_{T}(T)=(\sigma_S \sqrt{2\pi})^{-1}
e^{\frac{1}{2}(\frac{T-\mu}{\sigma_S})^2}$ with $\mu\simeq 25 \mu K$ and
$\sigma_S \simeq 200 \mu K$. Starting with this temperature distribution 
$f_{T}(T)$, after taking $n$-pixel demotion, we can derive the distributions 
of $L_x$ and $L_y$,
\begin{equation}\label{v}
f_{comp}(L_x)=\frac{1}{\sigma_n\sqrt{2\pi}}e^{-\frac{L^2_x}{2\sigma^{2}_n}}
\end{equation}
where $\sigma^2_n$ is the variance for temperature distribution for $n$-pixel 
demoted pixels for example, $\sigma_{16}\simeq 104\mu K$. $L_y$ also has almost same distribution as equation (\ref{v}), so we use equation (\ref{v}) for both 
$L_x$ and $L_y$. This gives the probability distribution function for change of 
$x$-component, $\Delta_x \equiv L_x^{i} -L_x^{i+1}$,
\begin{equation}\label{w}
f_{\Delta_x}(\Delta_x)=\frac{1}{2\sigma_n\sqrt{\pi}}
e^{-\frac{\Delta^2_x}{4\sigma^2_n}}.
\end{equation}
and the same function for $y$-component. Since a pattern formed by moving 
string should have relatively constant height of step along the curve of 
pattern, we can impose a condition for connectedness as
\begin{equation}\label{x7}
|\Delta_x|<\Delta L_c,\:\: |\Delta_y|<\Delta L_c
\end{equation}
then, the probability that two adjacent vectors 
$\vec L_{i}$ and $\vec L_{i+1}$ meet these conditions is given by
\begin{eqnarray}
P_{i,i+1}&=&\left[\int^{\Delta L_c}_{-\Delta L_c}
f_{\Delta_x}(\Delta_x)d\Delta_x\right]^2 \nonumber\\
&=&\left[erf(\frac{\Delta L_c}{2\sigma_n})\right]^2 \label{x}
\end{eqnarray} 
where $\Delta L_c$ is the maximum value allowed for $\Delta_x$ and $\Delta_y$. 

We set another condition for connectedness 
\begin{equation}\label{x8}
\theta_{i,i+1}\equiv\cos^{-1}(\hat D_{i}\cdot\hat D_{i+1})<\theta_c
\end{equation}
for a sequence to avoid too sharp turns (Figure 7) where $\theta_{c}$ is the 
maximum angle allowed for $\theta_{i,i+1}$ to claim 
$\vec L_{i}$ and $\vec L_{i+1}$ are connected. Figure 8 and 9 show examples 
of patterns which comply the definition described in equations (\ref{x7}) and 
(\ref{x8}) found in the WMAP 1st-year data. Here, we set 
$\Delta L_c = \sigma_{16}=104\mu K$ and $\theta_c = \pi/6$.
\begin{figure}[t]
\setlength{\unitlength}{1cm}
\begin{center}
\begin{picture}(5,2.8)
\put(0,0){\line(1,0){5}}
\put(0,2.8){\line(1,0){5}}
\put(0,0){\line(0,1){2.8}}
\put(5,0){\line(0,1){2.8}}
\put(1.2,1.4){$\ldots\ldots\ldots\ldots$}
\put(1.2,1.4){$\ldots$}
\put(1,1.4){\vector(4,1){1.5}}
\put(1,1.4){\vector(1,1){0.7}}
\put(3,1.4){\vector(2,-1){1.5}}
\put(3,1.4){\vector(1,0){1.0}}
\put(4,1.4){$\ldots\ldots$}
\put(1.8,2.2){\makebox(0,0){.}}
\put(1.9,2.3){\makebox(0,0){.}}
\put(2,2.4){\makebox(0,0){.}}
\put(0.9,1.3){\makebox(0,0){.}}
\put(0.8,1.2){\makebox(0,0){.}}
\put(0.7,1.1){\makebox(0,0){.}}
\put(0.6,1.0){\makebox(0,0){.}}
\put(0.5,0.9){\makebox(0,0){.}}
\put(0.4,0.8){\makebox(0,0){.}}
\qbezier(1.45,1.85)(1.6,1.70)(1.62,1.4)
\put(1,1.4){\circle*{0.1}}
\put(3,1.4){\circle*{0.1}}
\put(0.8,1.9){\makebox(0,0){$\hat D_{i}$}}
\put(2.7,2.1){\makebox(0,0){$\vec L_i$}}
\put(3.7,1.8){\makebox(0,0){$\hat D_{i+1}$}}
\put(1.7,1.1){\makebox(0,0){$\theta_{i,i+1}$}}
\put(3.5,0.7){\makebox(0,0){$\vec L_{i+1}$}}
\end{picture}
\caption{Relative angles defined for connectedness. $\hat D_{i}$ is a 
unit vector along the line that connects $ \vec L_{i-1}$ and $\vec L_i$. 
$\theta_{i,i+1}$ is defined by the angle between $\hat D_{i}$ and 
$\hat D_{i+1}$, $\theta_{i,i+1}\equiv\cos^{-1}(\hat D_{i}\cdot\hat D_{i+1})$}.
\end{center}
\end{figure}
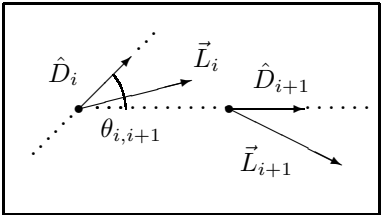
\begin{figure}[t]
\begin{center}
\epsscale{1.0}
\plotone{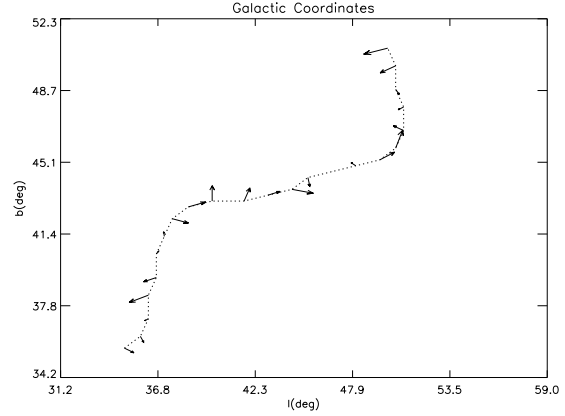}
\caption{A long temperature discontinuity pattern found at 
$(l=45^{\circ}$, $b=43^{\circ})$ in Galactic coordinates. The vector field is 
derived from 16-pixel demoted temperature sky map. This sequence has its 
maximum likelihood defined in the equation (\ref{x9}) at $\delta T=12\mu K$.}
\end{center}
\end{figure}
\begin{figure}[t]
\begin{center}
\epsscale{1.0}
\plotone{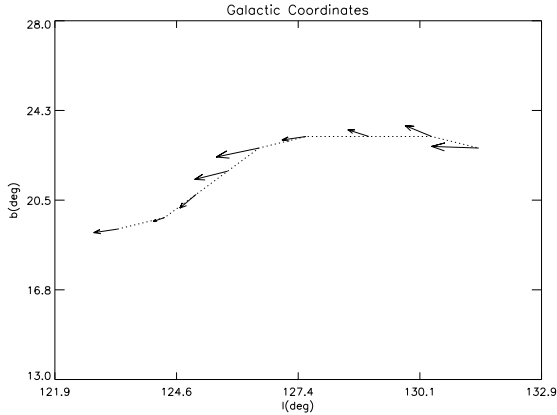}
\caption{A short pattern with 16-pixel demotion from WMAP 1st-year data 
located at $l=127^{\circ}$, $b=20^{\circ}$ in Galactic coordinates. 
$\delta T_{opt}$ for this pattern is $0\mu K$, i.e, no preferred bias is found 
which is against its apparent feature. This is because the temperature step 
vectors on the left-half of the curve have wrong direction relative to 
expected string velocity at this part while those on the right-half are 
aligned correctly and these two opposite contributions cancel out the bias.}
\end{center}
\end{figure}\\
(2) {\it Likelihood of sequence}\\
Given that a sequence of temperature steps defined in the previous paragraph, 
we can estimate its likelihood for a signal due to a moving string. If a long 
temperature step is formed solely by a moving string, the $\vec L$'s assigned 
on the curve should be tangent to the curve. After allowing contamination by 
noise, adiabatic fluctuation or by other possible sources, each $\vec L$ on the
curve will be off from local tangent. But if the contamination is not 
overwhelming, there should be still a bias seeded by string. We define the 
bias of a sequence with $N$-connected arrows quantitatively in terms of 
relative likelihood function as follows
\begin{eqnarray}
{\cal{L}}^{\left(N\right)}_{bias}(\delta T)&=&\prod^{N}_{i=1}
e^{-\left(\vec{L_i}-\phi_i\vec{\delta T_i}\right)^2/2\sigma^2_n}\label{x9}\\
 \phi_i &=&\left\{ \begin{array} {r@{\quad:\quad}l}
+1 & {\mathrm{right\!-\!handed\: curve}} \\ 0 & {\mathrm{straight\: line}} \\ 
-1 & {\mathrm{left\!-\!handed\: curve}}
\end{array} \right.\nonumber
\end{eqnarray}
where $\vec{\delta T_i}$ is a tangent vector at $i_{th}$ grid point on the curve
at which $\vec L_i$ is located and $|\vec{\delta T_i}|= \delta T$. 
$\phi_i$ is a phase factor defined at $i_{\mathrm{th}}$ grid point on the 
curve to give correct direction of string velocity on the curve. If the 
sequence turns right(left) locally at $i_{\mathrm{th}}$ grid, then 
$\phi_i=+1(-1)$ and it is zero when the sequence is straight at that point.
On both the head and tail of a sequence, the phase factors are set to zero. 
This is an approximate prescription to describe realistic model of string 
motion. When $\vec{L_i}$'s are perfectly coincident with $\vec{\delta T_i}$, 
then ${\cal{L}}^{\left(N\right)}_{bias}(\delta T)$ become 1 and if 
$\vec{L_i}$'s are off by either direction or magnitude or both from 
$\vec{\delta T_i}$'s, then ${\cal{L}}^{\left(N\right)}_{bias}(\delta T)$ 
decays exponentially. So, if there is nonzero $\delta T$ that gives maximum 
${\cal{L}}^{\left(N\right)}_{bias}(\delta T)$, then it is relatively more 
likely that there is a constant temperature step with height $\delta T$ 
imbedded in the sequence. Figure 10 shows the comparison of results between 
actual data (WMAP 1st-year data) and a simulated white noise. We can estimate 
from the curve the height of temperature discontinuity in equation (\ref{a}),
\begin{equation}\label{z2}
\delta T \lesssim 42\mu K ,\:\: 95\% {\mathrm {CL}}
\end{equation}
which is equivalent to symmetry breaking scale 
$\eta_{SB} \sim 0.94\times 10^{16}$GeV. 
\begin{figure}[t]
\begin{center}
\epsscale{1.0}
\plotone{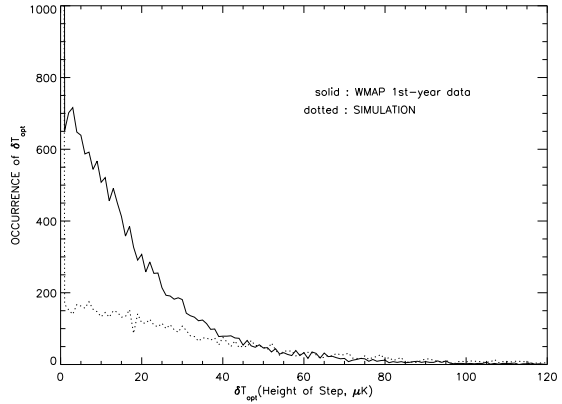}
\caption{Distribution of $\delta T_{opt}$ at which likelihood function defined 
in equation (\ref{x9}) becomes maximum for a sequence. Each data set takes 
16-pixel demotion. We included sequences which contain five or more arrows 
connected. There are many more sequences in the actual data than in the
simulation which has identical size, mean and variance but white noise.}
\end{center}
\end{figure}  
\section{Conclusion}
We have investigated WMAP 1st-year data to search directly or set limit on cosmic strings. For statistics of full sky map, we set the limit of contribution to
variance by the strings as $0<\sigma_{string}<22.5\muK$ which gives upper limit 
on deficit angle $0 < \delta = 8\pi G\mu /c^2 <0.82\times 10^{-5}$.
And we set up a model with a random distribution of strings with random
orientation with which the relation between $\sigma_{string}$ and height of
temperature step $8\pi G\mu \gamma \beta/c^2$ can be found.
A limit on deficit angle is obtained from the model of uniform distribution of
strings, $0 < \delta = 8\pi G\mu /c^2 <7.34\times 10^{-5}$. This corresponds 
to symmetry breaking energy scale $\eta_{SB}\sim m_{pl}\sqrt{\frac{G\mu}{c^2}}
\sim 2\times 10^{16}$GeV.

We developed a pattern search algorithm that can visualize the landscape of
CMB temperature variation of the sky. There were some fairly long temperature
rows but we didn't find any compelling pattern of cosmic strings predicted by
theory. Instead, by considering the distribution of heights of temperature 
discontinuities, we roughly estimated $0<\delta = 8\pi G\mu /c^2< 1.54 \times 
10^{-5}$ or equivalently $\eta_{SB} \sim 0.94\times 10^{16}$GeV.

The precision of data is yet to be refined and we expect WMAP 2nd-year data
will provide much better chance to pin down the effects of topological defects 
on cosmic microwave background radiation and at that point refined analysis 
will be appropriate.
\section{Acknowledgements}
This work was supported by the U.S Department of Energy under Contract 
No. DE-AC03-76SF00098 at LBNL Physics Division and Physics Department at 
University of California, Berkeley. Some of the results in this paper have 
been derived using the HEALPix\footnote{http://www.eso.org/science/healpix/} 
(G\'{o}rski, Hivon, and Wandelt 1999). 
We would like to thank E. Canudas, K. Howley, J. Lamoreaux, T. Watari and 
L. Zuniga for discussion and comments.


\begin{thebibliography}{99}
\bibitem[Jeannerot (2003)]{jeannerot} [1] R. Jeannerot, J. Rocher, M. Sakellariadou, astro-ph/0308134 (2003)
\bibitem{cmb1} [2] H.V. Peiris, et al., astro-ph/0302225 (2003)
\bibitem{cmb2} [3] L. Pogosian, S.-H. Henry Tye, I. Wasserman, M. Wyman,
   astro-ph/0304188 (2003)
\bibitem{sarangi2} [4] S. Sarangi and S-H.H. Tye, Phys. Lett. B536 (2002) 185, hep-th/0204074
\bibitem{Jones3} [5] N. Jones, H. Stoica and S.-H.H. Type hep-th/0303269
\bibitem{Vachaspati1986} [6] Vachaspati, T.  1986 'Gravitational effects of cosmic strings' Nucl. Phys. B277, 593.
                                                                                
\bibitem{Vilenkin1986b}[7] Vilenkin, A 1986 "Looking for Cosmic Strings', Nature 322, 613.
                                                                                
\bibitem{KaiserStebbins1984} [8] Kaiser, N. \& Stebbins, A. 1984 'Microwave anisotropy due to cosmic strings',  Nature 310, 391.
                                                                                
\bibitem{Gott1985}  [9] Gott, J.R. ``Gravitational Lensing Effects of Vacuum String: Exact Results'  Ap. J. 288, 422.
\bibitem{Bonometto2002} [10] S. Bonometto, V. Gorini, U. Moschella {\it Modern Cosmology} IOP(2002)
\bibitem{lamoureaux}[11] J. Lamoureaux, L.Zuniga, K. Howley, G. F. Smoot,
``A New Technique for the Detection of Cosmic Strings in the GOODS Data'',
in preparation (2004).
\bibitem{Urrestilla}[12] J. Urrestilla, A. Ach\a'ucarro, A. C Davis, 
hep-th/0402032
\bibitem{Watari}[13] Taizan Watari, T. Yanagida, hep-ph/0402125
\bibitem{Dasgupta}[14] Keshav Dasgupta, Jonathan P. Hsu, Renata Kallosh, 
Andrei Linde, Marco Zagermann, hep-th/0405247
\end{thebibliography}
\end{document}